\documentclass[english,nofootinbib,superscriptaddress]{revtex4}
\usepackage{lmodern}

\usepackage[T1]{fontenc}
\usepackage[latin9]{inputenc}
\usepackage[letterpaper]{geometry}
\usepackage{amsmath}
\usepackage{amssymb}
\usepackage{bbm}
\usepackage{babel}
\usepackage{graphicx}

\begin{document}

\title{Modeling transverse relative locality}

\author{Giovanni AMELINO-CAMELIA}
\affiliation{\footnotesize{Dipartimento di Fisica, Universit\`a di Roma ``La Sapienza", P.le A. Moro 2, 00185 Roma, EU}}
\affiliation{\footnotesize{INFN, Sez.~Roma1, P.le A. Moro 2, 00185 Roma, EU}}
\author{Leonardo BARCAROLI}
\affiliation{\footnotesize{Dipartimento di Fisica, Universit\`a di Roma ``La Sapienza", P.le A. Moro 2, 00185 Roma, EU}}
\author{Niccol\'o LORET}
\affiliation{\footnotesize{Dipartimento di Fisica, Universit\`a di Roma ``La Sapienza", P.le A. Moro 2, 00185 Roma, EU}}
\affiliation{\footnotesize{INFN, Sez.~Roma1, P.le A. Moro 2, 00185 Roma, EU}}

\begin{abstract}
We investigate some aspects of relativistic classical
theories with ``relative locality",
in which pairs of events established to be coincident by nearby observers
may be described as non-coincident by distant observers.
While previous studies focused mainly on the case of longitudinal relative locality,
where the effect occurs along the direction connecting the distant observer
to the events, we here focus on transverse relative locality, in which instead the
effect is found in a direction orthogonal to the one
connecting the distant observer to the events.
Our findings suggest that, at least for theories of free particles
such as the one in arXiv:1006.2126,
 transverse relative locality
is as significant as longitudinal relative locality both conceptually and quantitatively.
And we observe that ``dual gravity lensing", first discussed in
arXiv:1103.5626, can be viewed as one of two components of transverse relative locality.
We also speculate about a type of spacetime noncommutativity for which
 transverse relative locality  could be particularly significant.\\

\end{abstract}

\maketitle

\section{Introduction}
The possibility of a novel ``relativity of locality" is being
proposed~\cite{prl,grf2nd} as a candidate
feature of quantum gravity.
When locality is relative processes are still local in the coordinatization
of spacetime by nearby observers, but they may appear to be nonlocal
in the coordinatization of spacetime by distant observers.
One can appreciate some of the main implications of this hypothesis
by looking, for example, at a particle decay $a \rightarrow b+c$
decomposed
into a disappearance event for particle $a$ and two particle-production events
for particles $b$ and $c$: with absolute locality the 3 events coincide
in the coordinatization of spacetime by any observer, but with relative locality
only ``nearby observers" (observers themselves ``local to" the coincidence of events)
describe the events as sharply coincident, whereas distant observers in general
describe them as not exactly coincident.
Of course, since so far all our observations are consistent with absolute locality,
it must be assumed that the characteristic scale
of the relativity of locality is very small, and the quantum-gravity
intuition that motivates these studies provides~\cite{prl,grf2nd} a natural
candidate for such a scale: the inverse of the Planck scale $M_p \sim 10^{28}eV$.

We shall here not dwell on the strength of the quantum gravity motivation
for studies of relative-locality, for which readers find detailed arguments
in Refs.~\cite{prl,grf2nd}. Our perspective is focused on investigating
relative locality
as a novel candidate relativistic feature, whose full understanding
we expect to surely require
 as much dedicated effort as was needed for
the Galilean relativity
of rest and the Einstenian relativity of simultaneity.
For this purpose we return here to the simple perspective
and formalization
of the first studies~\cite{whataboutbob,leeINERTIALlimit,arzkowaRelLoc}
which noticed that some known deformations of relativistic symmetries,
some of the ones studied within the ``doubly-special relativity"
research programme (see, {\it e.g.},
Refs.~\cite{gacdsr1,kowadsr,leedsrPRL,dsrnature,leedsrPRD,jurekDSRnew}),
could result in the property that coincidences of events established
by nearby observers might not appear as coincidences of events in
the coordinatization of spacetime by distant observers.
The formalization adopted in Refs.~\cite{whataboutbob,leeINERTIALlimit,arzkowaRelLoc}
has the limitation of being confined to the description of free
particles, which is evidently a severe limitation for what concerns
the development of realistic physical theories with relative locality.
This limitation
was removed by the more powerful formulation of relative locality,
including a description of interactions,
advocated in Refs.~\cite{prl,grf2nd}
(also see Refs.~\cite{leelaurentGRB,soccerball,flaviogiuliaRL,flaviojoseRL,anatomy}).
But for the characterization of certain relativistic issues,
such as the ones which are here
of our interest, the restriction to free particles is not
an important limitation, and the simplicity of the formalism proves
to be very advantageous.

Our objective here is to advance the understanding of relative locality
by introducing a distinction between longitudinal relative locality
and transverse relative locality:
with longitudinal relative
 locality coincidences of events established by nearby observers
are described by distant observers as events that are non-coincident along the
direction
connecting the observer to the events, whereas
with transverse relative locality
the distant observer describes the events as non-coincident along a
direction orthogonal to the direction
connecting observer to events.

All relative-locality studies produced so far mainly focused
on longitudinal relative locality.
A brief mention of a feature of transverse relative locality
is found in Ref.~\cite{whataboutbob}: the analysis in
Ref.~\cite{whataboutbob} nearly exclusively focused on the case
of coincidences of events established
by nearby observers which would be described by distantly boosted
observers as events that do not
coincide along the
direction
connecting the distantly boosted observer to the events
(a case of longitudinal relative locality), but
Ref.~\cite{whataboutbob}  also mentioned briefly
the possibility that for some
distantly boosted observers the lack of coincidence
could also be ``transverse", a lack of coincidence occurring in
a direction
orthogonal to the direction connecting the distantly boosted observer
to the events.
Another brief appearance of a manifestation of what we here label as transverse
relative locality can be found implicitly in parts of Ref.~\cite{leelaurentGRB},
specifically the parts of Ref.~\cite{leelaurentGRB}
that concerned some analyses of the formalism
of Refs.~\cite{prl,grf2nd}, for interacting particles,
providing evidence of ``dual-gravity lensing", intended as a manifestation
of relative locality such that particles on parallel propagation according
to some observers could be described
by other observers as propagating along different directions.
As discussed in some detail here below,
this ``dual-gravity lensing"\footnote{One can qualify this sort of
effects as ``dual-gravity lensing" in light of the thesis put forward
in Refs.~\cite{prl,grf2nd} which characterizes relative locality
as a manifestation of the, possibly curved, geometry of momentum space.
The standard gravitational lensing is caused by spacetime curvature, and
this relative-locality-induced ``lensing" can be attributed, in
light of Refs.~\cite{prl,grf2nd}, to the ``dual gravity" of momentum space.}
is intimately connected with transverse
relative locality.

We are here reporting a first dedicated study of transverse relative locality,
whose humble objectives are focused on building a few elements of intuition
on transverse relative locality, its possible dependence on different
scales in some applications of interest, and its comparison to longitudinal relative
locality. We shall in particular show that, at least within
 the confines of relativistic theories of free particles,
 all previously-obtained results for longitudinal relative locality
 have a clear (and no less significant) counterpart on the transverse-relative-locality
 side. Indeed in the next section we find that the doubly-special-relativity-based
 formalism that led, in Ref.~\cite{whataboutbob}, to the derivation
 of manifestations of longitudinal relative locality for some distantly boosted
 observers also produces transverse relative locality for some other
 distantly boosted  observers, and the magnitude of the two classes of effects
 is comparable.
Then in Section III we also expose some transverse-relative-locality
 coordinate artifacts that result from adopting spacetime-noncommutativity-inspired
 phase-space constructions, just like
in  phase-space constructions inspired
 by $\kappa$-Minkowski noncommutativity~\cite{lukieIW,majidruegg,kpoinap}
 it was recently established~\cite{kappabob}
 that longitudinal-relative-locality
 coordinate artifacts are present.
 So the evidence we here provide suggests, however preliminarily,
 that longitudinal and transverse relative locality really need to be considered
 as equally meaningful aspects of relative locality.

We work throughout in 2+1 spacetime dimensions. Evidently transverse relative locality
cannot be present in 1+1-dimensional theories, and on the other hand all
the  features
of transverse relative locality
we are here interested in  are fully describable within a 2+1-dimensional setup
(the generalization to a D+1-dimensional analysis is only modestly cumbersome,
but adds nothing to the concepts and results here of interest).

\section{Transverse relative locality from distant boosts}
Our first task is to expose the presence of transverse relative locality
and of the associated dual-gravity lensing
within the phase-space setup introduced in Ref.~\cite{whataboutbob},
which was there used instead mainly to characterize longitudinal relative locality.
So we follow Ref.~\cite{whataboutbob} in introducing its three main ingredients:
(I) ordinarily trivial Poisson brackets for the spacetime coordinates
($\{ x_j , t \}=0~,~~\{ x_j , x_k \}=0$); (II)
a completely standard description of the generators of space ($P_j$)
and time ($\Omega$) translations
\begin{eqnarray}
& \{\Omega,t\}=1, \quad \{\Omega,x_i\}=0~, \nonumber\\
& \{P_i,t\}=0, \quad \{P_i,x_j\}=-\delta_{ij} ~,\nonumber\\
& \{P_i,P_j\}=0, \quad \{P_i , \Omega\}=0~;
\end{eqnarray}
and of the generator of rotation
$$\{R , x_i \} = \epsilon_{ij}x_j~,\{R , t \} = 0~,$$
$$\{R , P_i \} = \epsilon_{ij}P_j~,\{R , \Omega \} = 0~,$$
(III) but an unconventional description,
with deformation parameters $\alpha,\beta,\gamma$, of the Poisson brackets between
boost generators and generators of spacetime translations
\begin{eqnarray}
 \left\{ \mathcal{N}_i, \Omega \right\} &=& P_i-\alpha\ell\Omega P_i\label{poinc1}\\
\left\{ \mathcal{N}_i, P_j \right\} &=& \Omega\delta_{ij}+\ell\left((1+\gamma-\alpha)\Omega^2+\beta \vec{P}^2\right)\delta_{ij}-\ell\left(\gamma+\beta-\frac{1}{2}\right)P_iP_j~,\label{poinc2}
\end{eqnarray}
which in particular leads to~\cite{whataboutbob} the following
one-parameter family
of on-shell relations
\begin{equation}
\mathcal{C}_\ell = \Omega^2 - \vec{P}^2 +\ell( 2\gamma \Omega^3
+ (1-2 \gamma) \Omega \vec{P}^2 ),
\label{casimir}
\end{equation}

It is easy to verify~\cite{whataboutbob}
that all Jacobi identities are satisfied by these choices of Poisson brackets,
and readers familiar with the doubly-special-relativity
literature~\cite{gacdsr1,kowadsr,leedsrPRL,dsrnature,leedsrPRD,jurekDSRnew}
will recognize a rather standard DSR-type choice of boost generators.

For what concerns the derivation of
the equations of motion (worldlines)
in such a setup one
 can of course use~\cite{whataboutbob,jurekvelISOne,mignemi}
the on-shell relation as Hamiltonian
 of evolution  in an auxiliary worldline parameter $\tau$.
 Evidently the momenta are conserved on the worldlines
 (since $\{\mathcal{C}_\ell,P_j\} = 0 = \{\mathcal{C}_\ell, \Omega\}$).
 And one finds
\begin{eqnarray}
&\dot{t} \equiv \frac{\partial t}{\partial \tau} &
= \{\mathcal{C}_\ell,t\}= 2 \Omega + \ell (6 \gamma \Omega^2 +(1-2 \gamma)\vec{P}^2),\\
&\dot{x_i}
\equiv \frac{\partial x}{\partial \tau}
&=\{\mathcal{C}_\ell,x_i\}=2 P_i (1-\ell (1-2\gamma)\Omega)~,
\end{eqnarray}
 from which in particular one obtains that for massless particles ($\mathcal{C}_\ell=0$)
the worldlines are governed by
\begin{equation}
(x-x_0)_i = (1-\ell |\vec{P}|) \frac{P_i}{|\vec{P}|} \, (t-t_0)~.\label{wordli}
\end{equation}
The fact that the speed of massless particles here depends on
momentum\footnote{The coefficients of the terms $\ell \Omega^3$
and  $\ell \Omega \vec{P}^2$ in $C_\ell$ were arranged in Ref.~\cite{whataboutbob}
just so that this speed law for massless particles, $1-\ell |p|$,
would be produced. This is how from the more general two-parameter
case $\mathcal{C}_\ell = \Omega^2 - \vec{P}^2 +\ell( \gamma' \Omega^3
+ \gamma" \Omega \vec{P}^2)$
one arrives at the one-parameter case considered here and in Ref.~\cite{whataboutbob}:
$\mathcal{C}_\ell = \Omega^2 - \vec{P}^2 +\ell( 2\gamma \Omega^3
+ (1-2 \gamma) \Omega \vec{P}^2)$.}
is the main intriguing feature of this relativistic framework,
and was the subject of several investigations
(see, {\it e.g.}, Refs.~\cite{jurekvelISOne,gacMandaniciDANDREA,mignemiVEL,ghoshVEL}),
including some which established the presence of longitudinal relative
locality~\cite{whataboutbob,leeINERTIALlimit,arzkowaRelLoc,kappabob}.

We shall here expose the presence in this framework
of also transverse relative locality through a very explicit analysis.
This analysis is centered on the properties of a coincidence of events,
local to an observer Alice ({\it i.e.} occurring in the spacetime
origin of Alice's reference
frame), with all events in the coincidence being events of emission
of a massless particle, all propagating in the same direction (but with
different momenta).
The transverse relativity of locality will be evident
once we establish how such a coincidence of events at Alice is described by a distant
boosted observer, and specifically an observer who is at some distance from Alice
along the direction of propagation of the particles and boosted
in a direction orthogonal to the direction
of propagation of the particles.

In preparation for this analysis let us first establish how a single
massless particle emitted in Alice's origin and propagating along its $x_1$ axis
is described by an observer Bob, translated with respect to Alice along the $x_1$
direction, and by an observer Camilla, purely boosted with respect to Bob
along the direction $x_2$.
Evidently, in light of the equations for worldlines of massless
 particles derived above, according to Alice such a particle has worldline
\begin{eqnarray}
x^A_1(t^A)=(1-\ell p)t^A  ~,~~~
x^A_2(t^A)=0  ~\label{xA2}
\end{eqnarray}

In order to obtain Bob's description of this same worldline we must
use the translation generators:
\begin{eqnarray}
&\mathcal{T}_{a,a,0}\rhd t= t+a\{P_1,t\}-a\{\Omega,t\}=t-a \nonumber\\
 &\mathcal{T}_{a,a,0}\rhd x_1=x_1+a \{P_1,x_1\}-a\{\Omega,x_1\}=x_1-a \nonumber\\
 &\mathcal{T}_{a,a,0}\rhd x_2=x_2~,~~~
 \mathcal{T}_{a,a,0}\rhd P_i= P_i~,~~~ \mathcal{T}_{a,a,0}\rhd \Omega=\Omega,
 \nonumber
\end{eqnarray}
where, for definiteness and simplicity, we specialized to the case of
an observer Bob whose spacetime origin is distant
from Alice's spacetime origin just the right amount
to detect in his origin a massless soft ($\ell p \simeq 0$) particle emitted
from Alice's origin (so the translation is $\mathcal{T}_{a,a,0}$ with
 parameters $ a_t = a,~a_1 = a,~a_2 = 0$).
We therefore find that Alice's worldline (\ref{xA2})
is described by Bob as follows
\begin{eqnarray}
 x^B_1(t)= - a +(1-\ell p)(t^B+a) ~,~~~
 x^B_2(t)=0  ~\label{xB2}
\end{eqnarray}

We are now ready for the final step of our planned analysis of the massless particle
of generic momentum $p$ emitted in Alice's origin along Alice's $x_1$ axis, {\it i.e.}
we can now perform the DSR-deformed boost along the $x_2$ direction to obtain the
description of that particle according to observer Camilla.
 For that we can rely on
the representation of the (3-parameter family of) boosts given
in Eq.~(\ref{boosts}) which was already derived in Ref.~\cite{whataboutbob}:
\begin{equation}
\mathcal{N}_i =   x_i \Omega-t P_i + \ell \left(\alpha t \Omega P_i + x_i\left(\beta \vec{P}^2  + (1+\gamma-\alpha)  \Omega^2\right) - \left(\gamma+\beta -\frac{1}{2}\right) x_k P^k P_j\right)~.
\label{boosts}
\end{equation}

In particular this leads to the following action of the boosts on coordinates:
\begin{eqnarray}
\left\{ \mathcal{N}_i, t \right\} &=& x_i+\ell\left(2(1+\gamma-\alpha)x_i\Omega+\alpha tP_i\right)\label{boostedti}\\
\left\{ \mathcal{N}_i, x_j \right\} &=& t\delta_{ij}-\alpha\ell t\Omega\delta_{ij}+\ell \left(\gamma+\beta -\frac{1}{2}\right) \left((x_k P^k)\delta_{ij}+x_j P_i\right)- 2 \beta \ell x_i P_j.\label{boostedics}
\end{eqnarray}

Specializing these formulae to the case of a boost purely in the $x_2$
direction,
and acting with it on
the worldline (\ref{xB2})
we arrive at the sought Camilla description:
\begin{eqnarray}
&x_1^C(t^C)= -a+(1-\ell p)(t^C+a)\label{xC1}\\
&x_2^C(t^C)=-\xi_2 a +\xi_2 a \left(\alpha-\beta-\gamma+\frac{1}{2}\right)\ell p+\xi_2\left(1-\left(\alpha-\beta-\gamma+\frac{1}{2}\right)\ell p\right)(t^C+a)\label{xC2}
\end{eqnarray}
where $\xi_2$ is the boost parameter for the transformation from Bob
to Camilla, which is a pure boost along the $x^2$ direction.\\
Some indirect manifestations of transverse relative locality are already visible
looking at this single worldline.
In particular, by eliminating $t^C$ one obtains the projection
of the worldline in the $x_1^C , x_2^C$ plane
\begin{eqnarray}
x_2^C(x_1^C)=\xi_2\left(1-\left(\alpha-\beta-\gamma-\frac{1}{2}\right)\ell p\right)x_1^C+\ell\xi_2 a p
\end{eqnarray}
which has some remarkable properties.\\
We notice two main features that characterize this result with respect
to the corresponding result that applies in
the special-relativity limit ($\ell \rightarrow 0$):\\
\indent (I) when distances of order $\ell\xi_2 a p$ are
 within the reach of available experimental
sensitivities it will be appreciated
 that the worldline does not cross Camilla's spatial origin, a feature
we shall find convenient to label as ``shift";
\\
\indent (II) when $\ell\xi_2 p$ is within the reach of available angular resolutions
(and $\alpha-\beta-\gamma-\frac{1}{2} \neq 0$)
the angle in the $x_1,x_2$ plane by which Camilla sees the arrival of the particle
is momentum dependent, which is the
mentioned ``dual-gravity lensing".

None of this in itself provides a direct manifestation of
relative locality, but as we shall see these
two features of ``shift" and ``dual-gravity lensing"
do play a role in the size of the transverse relative
locality effects. At least within
the framework we are here adopting, the effect of transverse relative locality
could be described as composed of these
two features, even in the very tangible sense that the magnitude
of the transverse relative locality
is obtained combining the magnitudes of the shift and
of the dual-gravity lensing.

In order to examine the relative locality itself of course we must
analyze contexts with distant coincidences of events, and
the result we obtained above
for a single massless particle emitted by Alice toward Bob and Camilla
is all that we shall need in order to characterize such coincidences
of events. Let us start by focusing
on the case of 3 wordlines of that type; specifically 3 massless particles
all emitted simultaneously
in Alice's origin toward Alice's $x_1$ axis,
but two of them\footnote{Our choice of considering two soft massless particles 
and one hard massless particle is somewhat redundant but helps us keep the 
presentation clearer. With only one soft particle, plus the hard particle,
one could already infer all the properties of the transverse relative locality
which we are going to discuss (in fact in Fig.~1 only one soft and one hard particle
are shown). By contemplating two soft particles we have the luxury of seeing explicitly
that coincidences of emission events of soft particles still behave with absolute 
locality, and this then renders more evident how the event of emission of a hard
particle behaves anomalously (with relative locality). Moreover, there is a ``relativist
tradition" of viewing an event as a crossing of two worldlines, and from that perspective
our 3 simultaneous emission events can be viewed as two independent crossing events: 
the crossing of two soft worldlines and the crossing of the hard worldline with one of the soft worldlines (it is of course irrelevant which one of the soft worldlines is taken into account for this).
From this traditional relativist perspective one would describe the relativity
of locality as the fact that for Alice the two crossings coincide whereas according
to the coordinates of distantly boosted Camilla they do not coincide.}
are ``soft" ({\it i.e.} their momenta, $p^{(1)}$ and $p^{(2)}$ are small enough
that $\ell p^{(1)}$ and $\ell p^{(2)}$ can be neglected; $\ell p^{(1)} \simeq 0 \simeq \ell p^{(2)}$) while the third one has ``hard" momentum $p^{(3)}$ and we keep track
of terms with factors $\ell p^{(3)}$.
So we have set up a coincidence of emission events
established by the observer Alice, local to the coincidence,
and with the work done above we can establish immediately how the
relevant 3 worldlines are described by Camilla:
\begin{eqnarray}
x_1^{(1)C}(t^C)= t^C ~,~~~
x_2^{(1)C}(t^C)=\xi_2 t^C\label{xCa}
\end{eqnarray}
\begin{eqnarray}
x_1^{(2)C}(t^C)= t^C ~,~~~
x_2^{(2)C}(t^C)=\xi_2 t^C\label{xCb}
\end{eqnarray}
\begin{eqnarray}
&x_1^{(3)C}(t^C)= -a+(1-\ell p^{(3)})(t^C+a)\nonumber\\
&x_2^{(3)C}(t^C)=-\xi_2 a
+\xi_2 a \left(\left(\alpha-\beta-\gamma+\frac{1}{2}\right)\ell p^{(3)} \right)+\xi_2\left(1-\left(\alpha-\beta-\gamma+\frac{1}{2}\right)\ell p^{(3)} \right)(t^C+a)\label{xCc}
\end{eqnarray}
According to Camilla's coordinatization the wordlines
of the two soft particles (for which the $\ell$-deformation is not felt)
of course cross at $\{ -a, -\xi_2a, -a \}$, just as they would in an ordinarily special-relativistic
theory. But (see Fig.~1) the hard worldline
does not go through $\{ -a, -\xi_2a, -a  \}$ and notably when $t^C=-a$ the hard particle
has coordinates
  $\{ -a, -\xi_2 a (1-(\alpha -\beta - \gamma +\frac{1}{2})\ell p^{(3)}), -a\}$, from which we establish a transverse
relative locality
of $|\Delta x_2^C| = (\alpha -\beta - \gamma +\frac{1}{2}) \xi_2 \ell a p^{(3)}$.
Notice that (as also shown in Fig.~1) this amount of transverse relative locality can be described
as a combination of the term we labeled ``shift" and of a contribution
proportional to the term we labeled ``dual gravity lensing".



\begin{figure}[ht!]
\begin{center}
\includegraphics[scale=0.25]{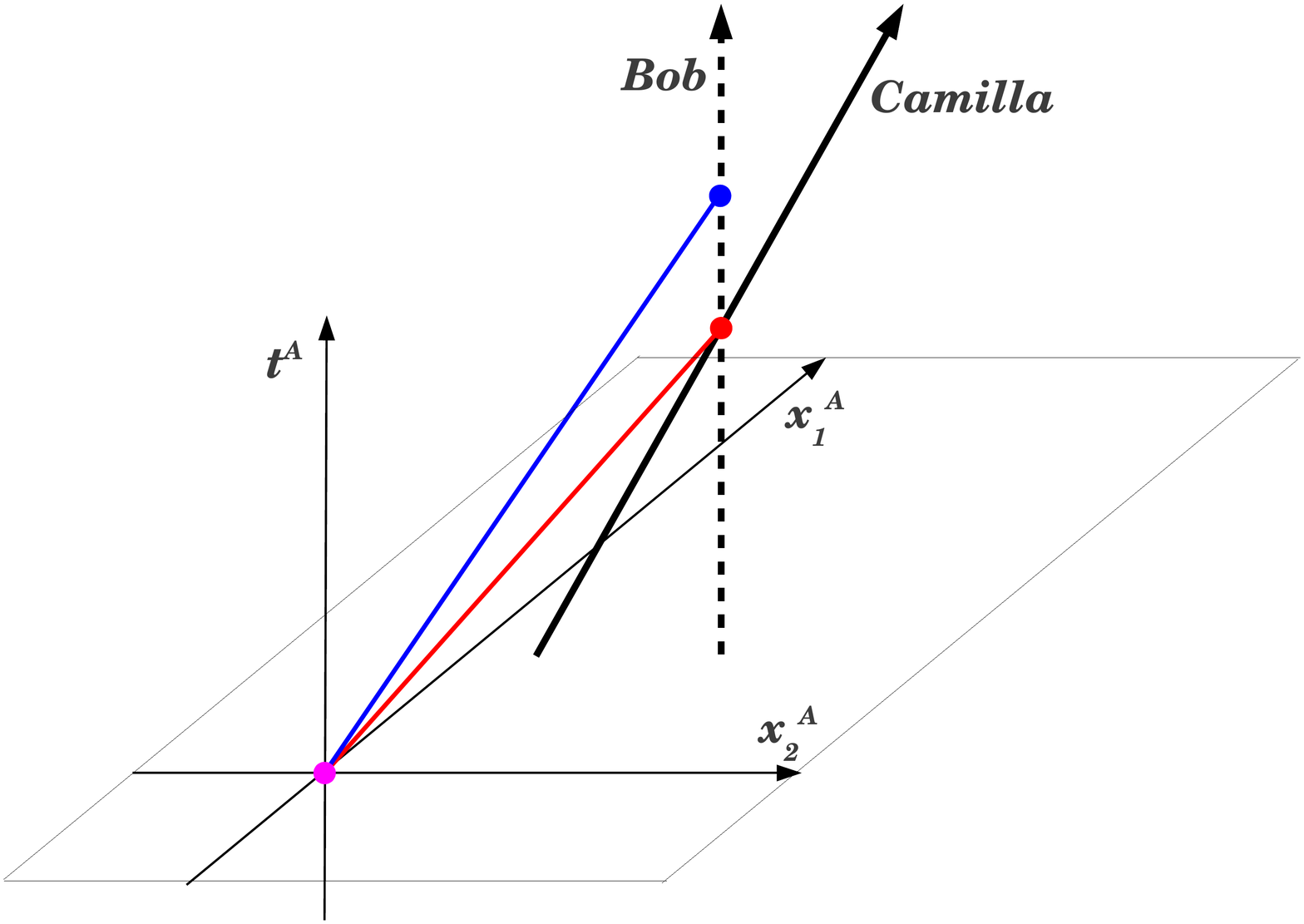}
\includegraphics[scale=0.25]{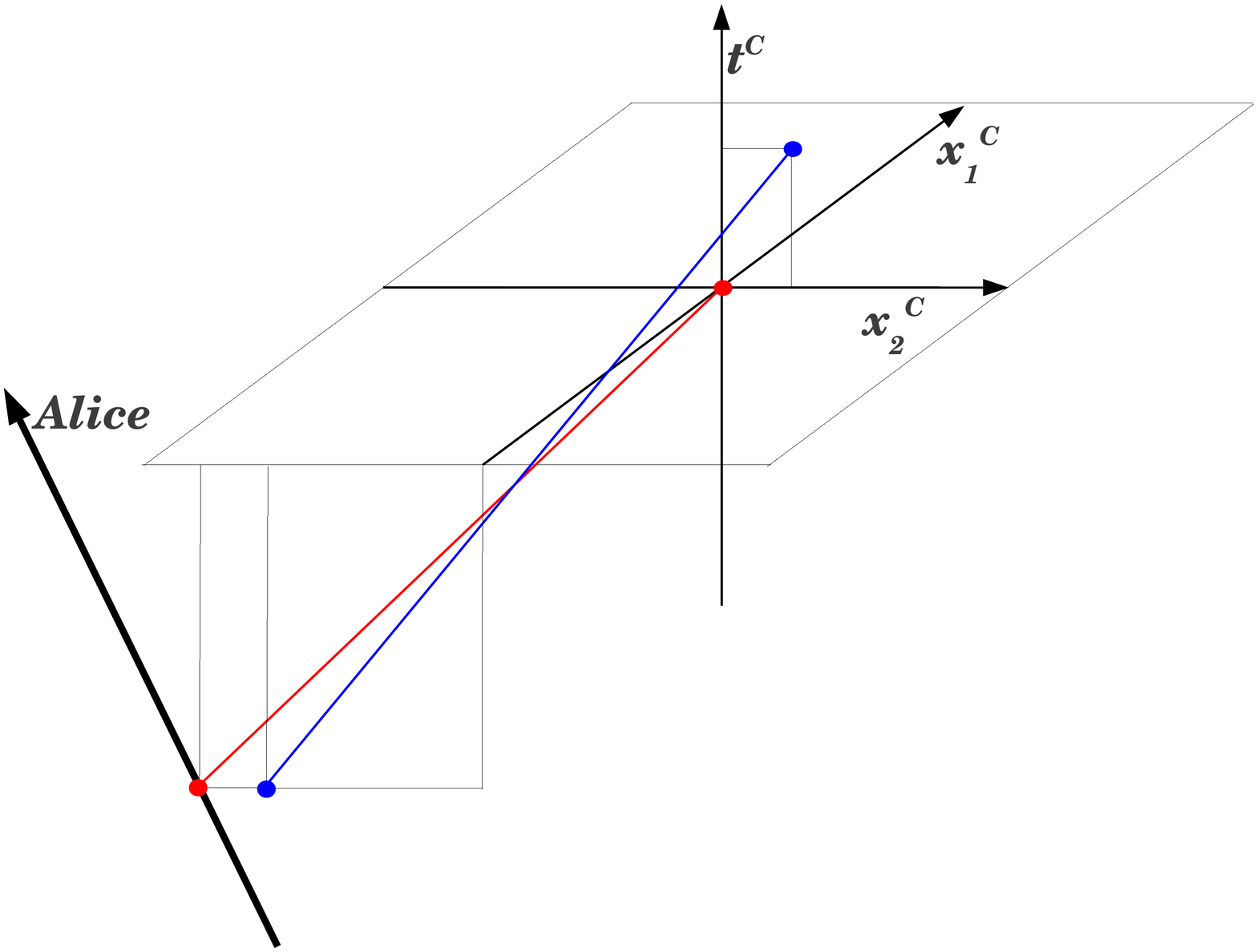}
\includegraphics[scale=0.23]{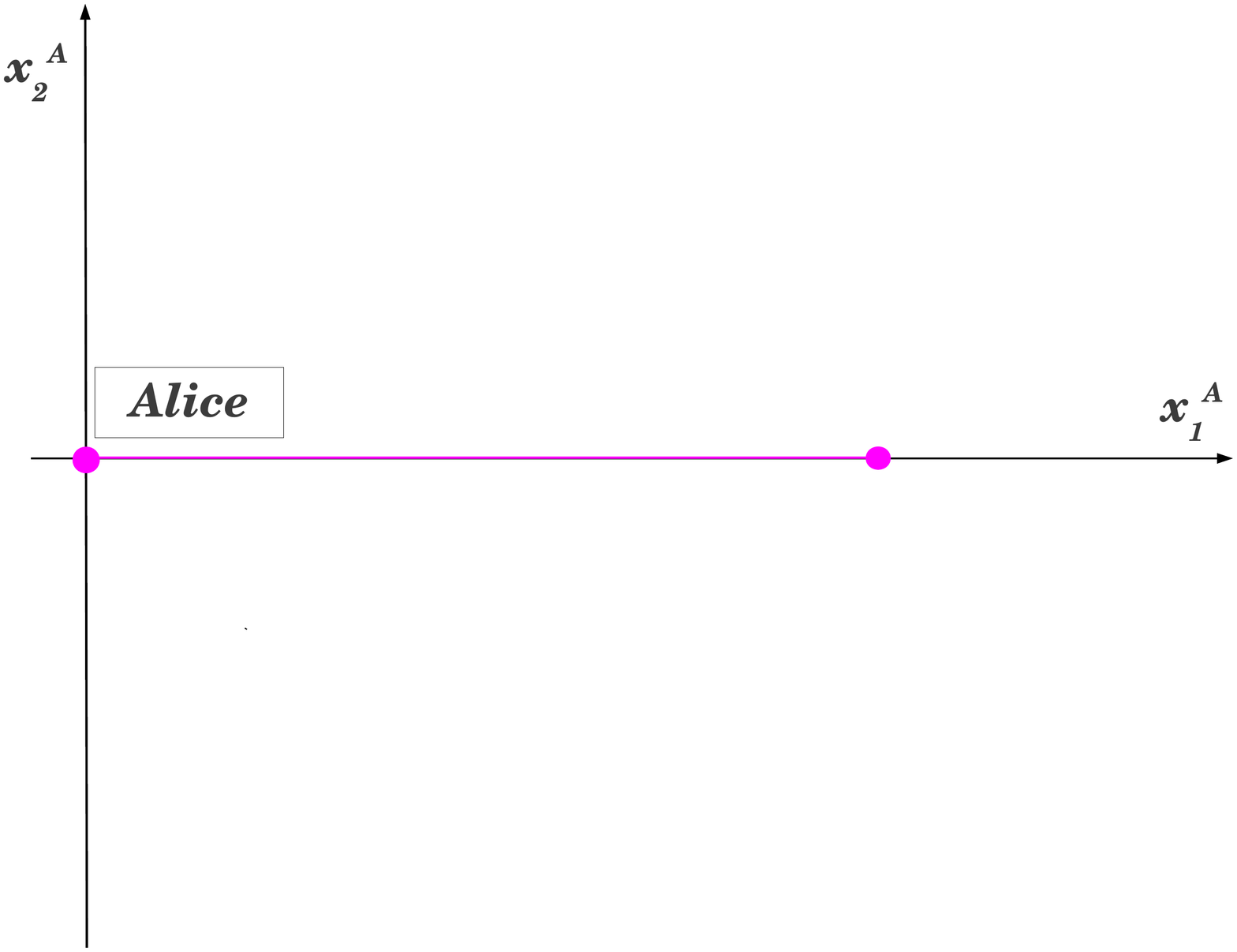}
\includegraphics[scale=0.23]{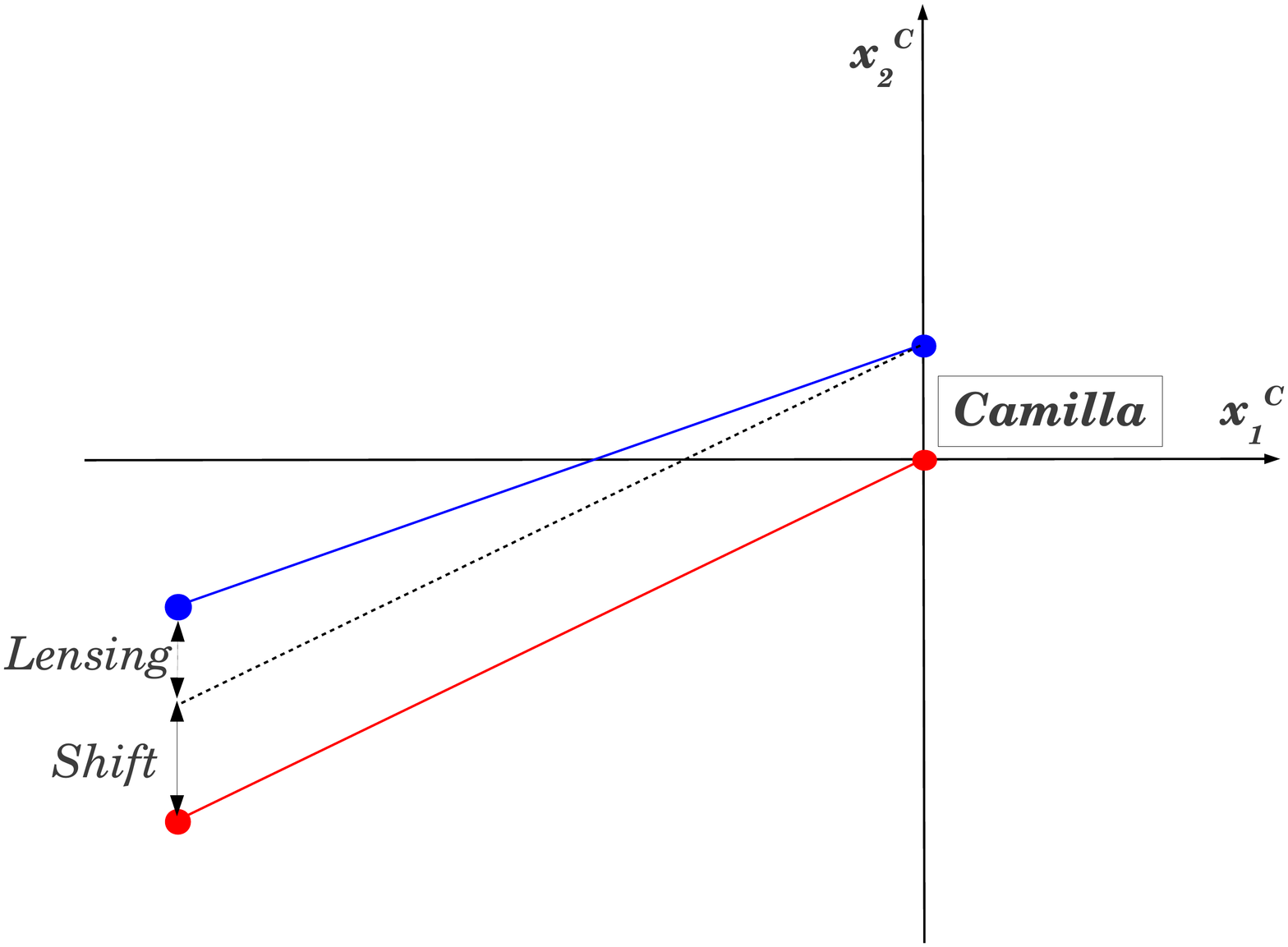}
\caption{3D worldlines
(top panels) and their 2D
spatial projection (bottom panels) for a soft and a hard (respectively red and blue; violet when coincident) massless particles emitted simultaneously at Alice
toward Bob and Camilla. Alice's viewpoint is shown in the left panels.
In Camilla's coordinatization (right panels)
the emissions are not coincident, and there is transverse
relative locality, with contributions
from ``shift" an ``dual-gravity lensing".}
\end{center}
\end{figure}

\newpage


So we have given a crisp characterization of how a distant coincidence of events,
the crossings of the 3 worldlines $p^{(1)}$,$p^{(2)}$,$p^{(3)}$ at Alice,
produces for a distantly boosted observer, Camilla, some transverse relative
locality, which in general also involves ``shift" and ``dual-gravity lensing".

It is interesting to check how the dual-gravity lensing depends on the momenta
of the particles, and for this purpose it is useful to contemplate a fourth
massless particle, with momentum $p^{(4)}$, again  emitted at Alice's origin
simultaneously
with the other 3 massless particles and emitted toward Alice's $x_1$ direction.
We take that $p^{(4)}$ is also ``hard", so that both terms with factors in $\ell p^{(4)}$
and in $\ell p^{(3)}$ should be taken into account.
For this fourth worldline Camilla's description evidently is
\begin{eqnarray}
&x_1^{(4)C}(t^C)= -a+(1-\ell p^{(4)})(t^C+a)\nonumber\\
&x_2^{(4)C}(t^C)=-\xi_2 a
+\xi_2 a \left(\left(\alpha-\beta-\gamma+\frac{1}{2}\right)\ell p^{(4)} \right)+\xi_2\left(1-\left(\alpha-\beta-\gamma+\frac{1}{2}\right)\ell p^{(4)} \right)(t^C+a)\label{xCd}
\end{eqnarray}

Comparing these with (\ref{xCc})
we see that in the $(x_1,x_2)$ plane the two hard wordlines reach Camilla
from directions forming an angle
$$\theta = \arctan\left(\xi_2 (1-(\alpha-\beta-\gamma-\frac{1}{2}) \, \ell p^{(4)}) \right) - \arctan\left(\xi_2(1-(\alpha-\beta-\gamma-\frac{1}{2}) \, \ell p^{(3)}) \right)\simeq -\xi_2 (\alpha-\beta-\gamma-\frac{1}{2}) \, \ell (p^{(4)} - p^{(3)})
$$
where notably the angle depends linearly on
the difference of the momenta $p^{(4)} - p^{(3)}$.

Finally, let us contemplate a different situation, with only two such massless
particles, of momenta $p^{(s)}$ and $p^{(h)}$ propagating again along Alice's $x_1$
direction but emitted from Alice's {\underline{spatial}} origin
with just the right  difference of times of emission that
they reach Bob's {\underline{spacetime}} origin simultaneously.
Assuming $p^{(s)}$ is soft ($\ell p^{(s)}\simeq 0$) and $p^{(h)}$ is ``hard"
($\ell p^{(h)}\neq 0$)
one has that the worldlines for these two particles are, according to Alice,
\begin{eqnarray}
& x_1^{(h)A}(t^A)=(1-\ell p^{(h)})(t^A+\ell a p^{(h)})~,~~~
 x_2^{(h)A}(t^A)=0
\end{eqnarray}
\begin{eqnarray}
 &x_1^{(s)A}(t^A)=t^A~,~~~x_2^{(s)A}(t^A)=0
\end{eqnarray}
Also for this situation we are interested in Camilla's coordinatization.
We have here a coincidence of (detection) events at Bob,
so Camilla is purely boosted with respect to an observer who is local
to a coincidence of events.
Using again the results we derived above one easily finds that Camilla describes
these two worldlines as follows
\begin{eqnarray}
 &x_1^{(h)C}(t^C)=-a+(1-\ell p_1^{(h)A})(t^C+a+ \ell a p_1^{(h)A})\nonumber\\
&x_2^{(h)C}(t^C)=-\xi_2 a\left(1-\left(\alpha-\gamma-\beta-\frac{1}{2}\right)\ell p_1^{(h)A}\right)+\xi_2\left(1-\left(\alpha-\beta-\gamma+\frac{1}{2}\right)\ell p_1^{(h)A}\right)(t^C+a+\ell a p_1^{(h)A})\nonumber
\end{eqnarray}
\begin{eqnarray}
 &x_1^{(s)C}(t^C)=t^C~,~~~x_2^{(s)C}(t^C)=\xi_2 t^C \nonumber
\end{eqnarray}
This allows us to verify that, as expected~\cite{prl,whataboutbob,kappabob},
the coincidence of events in Bob's origin is also described
as a coincidence of events by Camilla (both Camilla and Bob are nearby observers
of a coincidence of events, so even in a relative-locality framework they
should, and they do, agree on such coincidences of events).

There is however something noteworthy
about directions of propagation and for which it is useful to
characterize the two worldlines
 on the $(x_1,x_2)$ plane:
\begin{eqnarray}
& x_2^{(h)C}(x_1^{(h)C}) = \xi_2 (1- (\alpha - \beta - \gamma - \frac{1}{2}) \ell  p^{(h)}) x_1^{(h)C}\\
& x_2^{(s)C}(x_1^{(s)C}) = \xi_2 x_1^{(s)C}
\end{eqnarray}
From this we see that,
 while as expected no relative locality is
seen by Camilla in such situations, the dual-gravity lensing survives
(also see Fig.~2):
the two worldlines emitted by Alice along parallel directions
 are detected by Camilla along directions forming an angle (if $\alpha-\beta-\gamma-\frac{1}{2} \neq 0$)
 $$\theta = \arctan\left(\xi_2 (1-(\alpha-\beta-\gamma-\frac{1}{2}) \, \ell p^{(h)}) \right) - \arctan(\xi_2) \simeq - \, \xi_2 \,(\alpha-\beta-\gamma-\frac{1}{2}) \, \ell p^{(h)}$$

\begin{figure}[h]
\begin{center}
\includegraphics[scale=0.24]{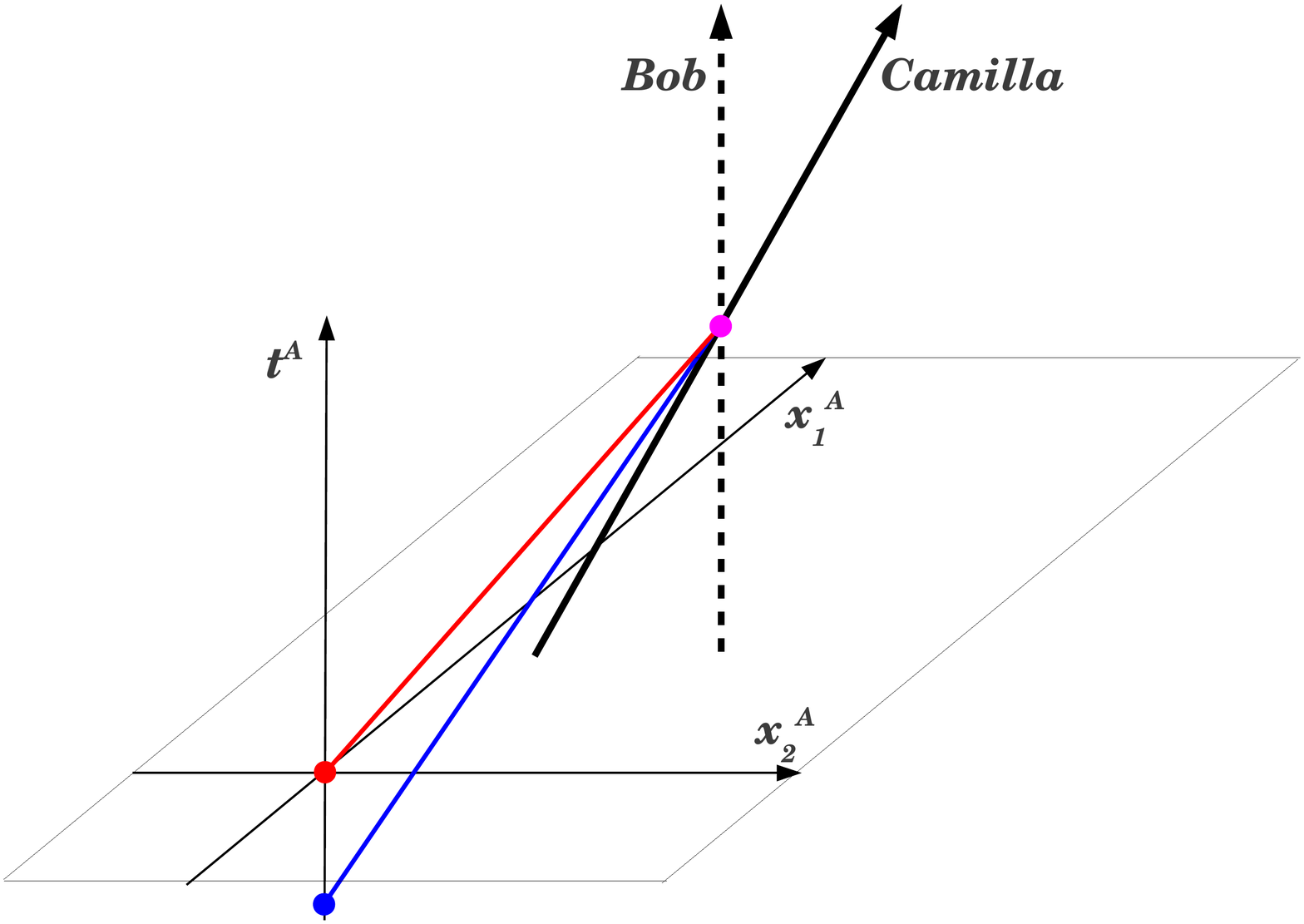}
\includegraphics[scale=0.24]{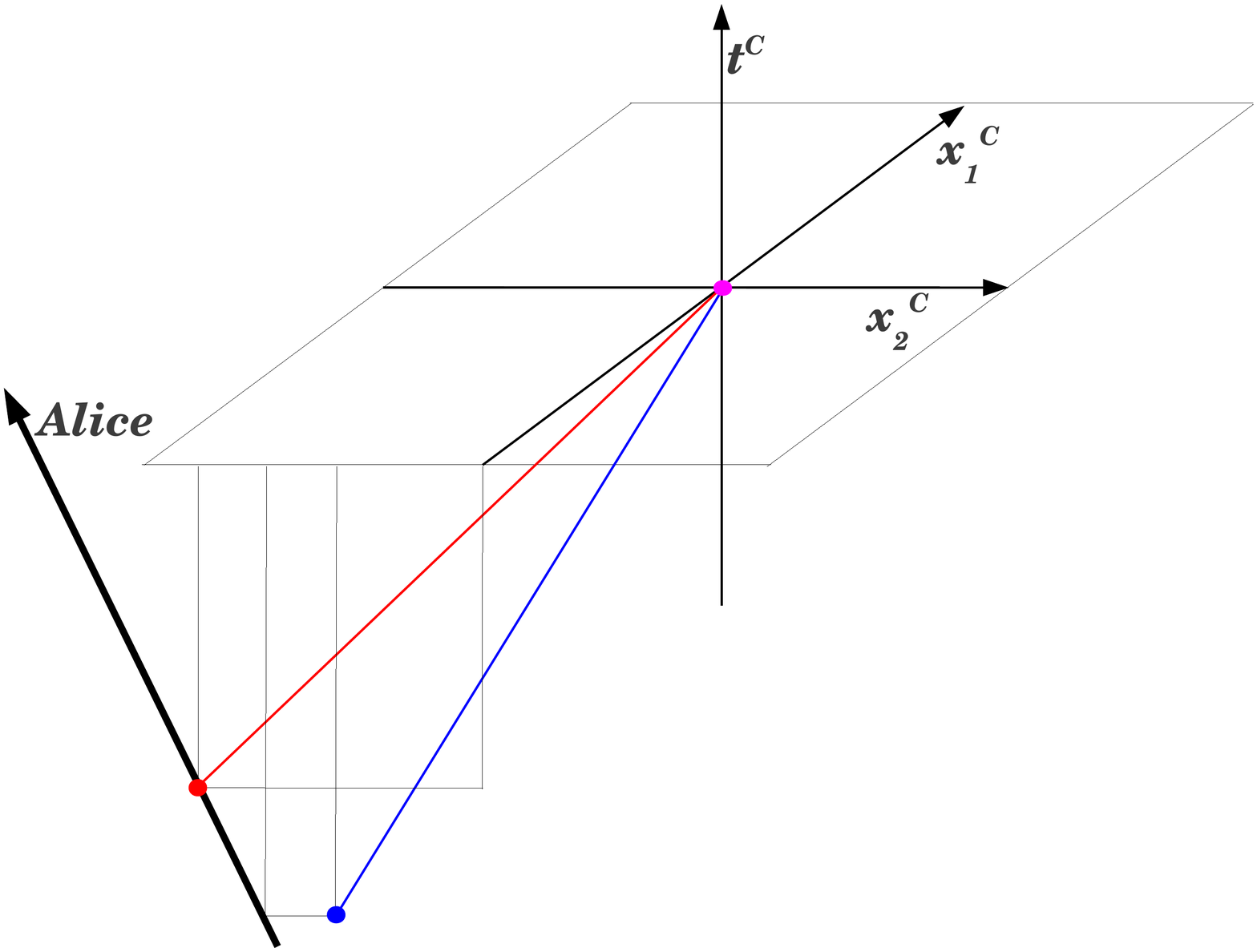}
\includegraphics[scale=0.22]{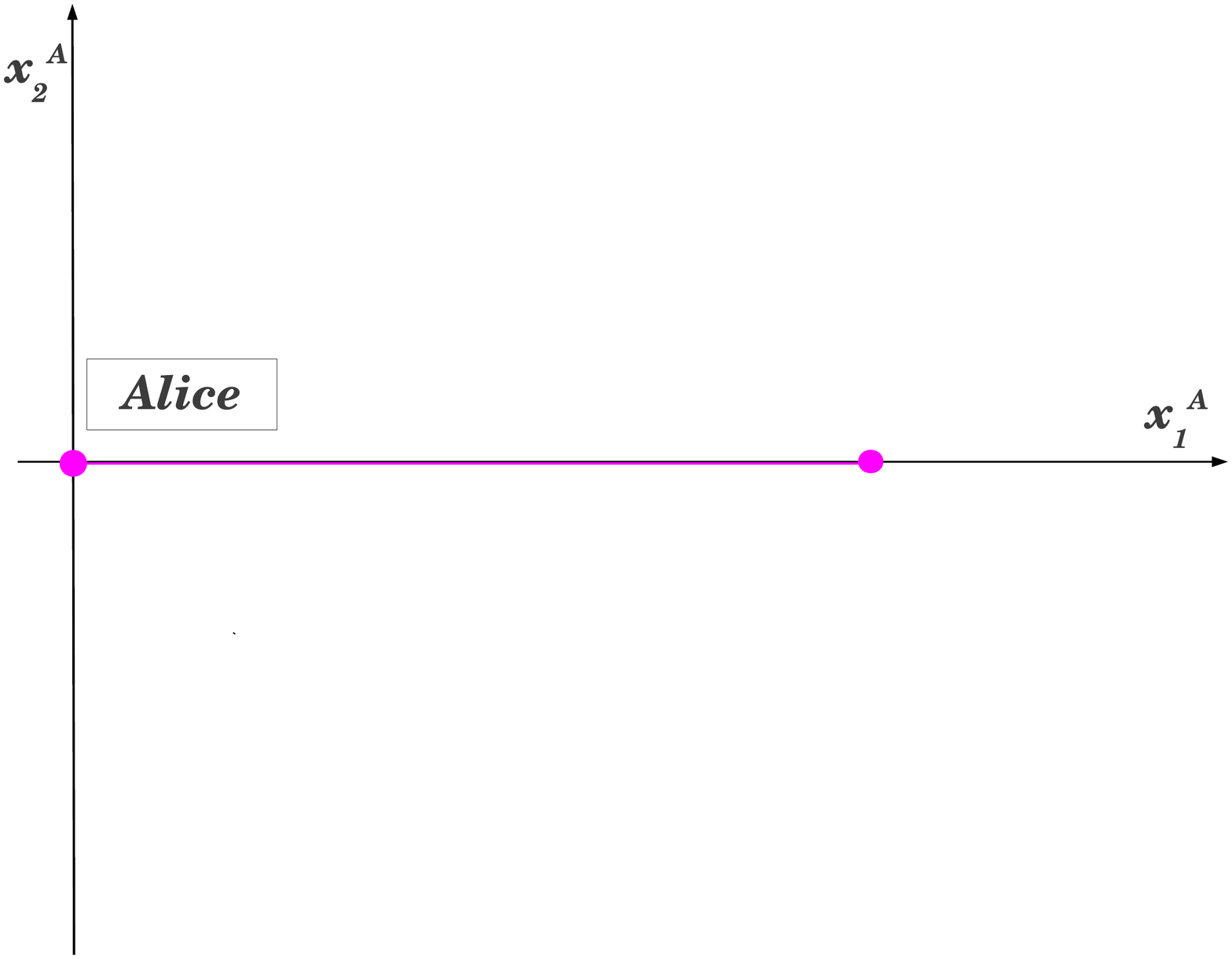}
\includegraphics[scale=0.22]{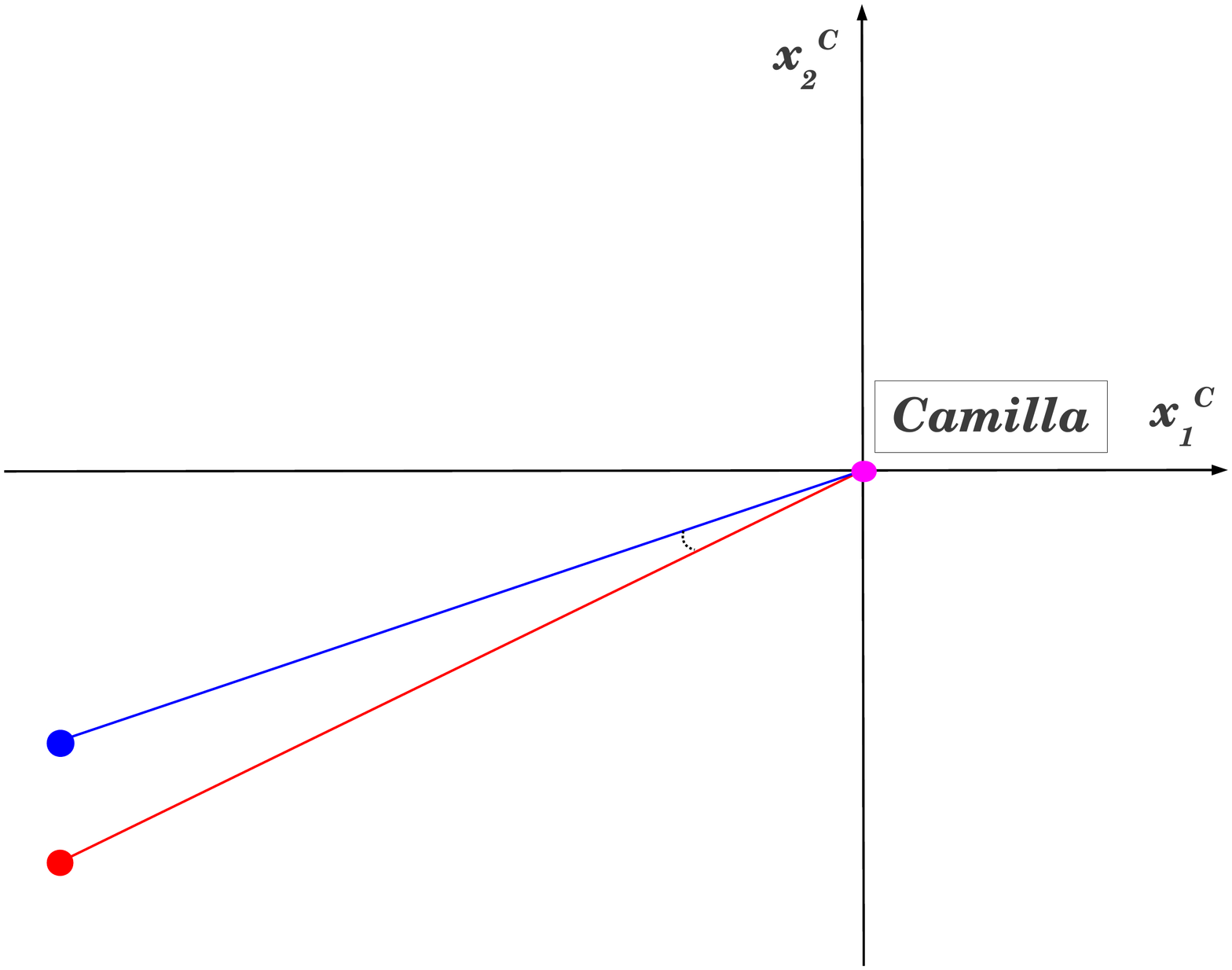}
\caption{Alice emits the soft (red) and the hard (blue) massless particles
from her spatial origin with a time-of-emission difference such
that they reach Bob simultaneously.
The simultaneous arrival of the two particles is also manifest in Camilla's coordinates.
But there still is some ``dual-gravity lensing":
whereas according to Alice's coordinatization
the particles are on parallel trajectories, according
to Camilla's coordinatization the particles are not on parallel trajectories.}
\end{center}
\end{figure}


\section{Noncommutativity-inspired transverse relative locality from pure translations}
In the previous section we established for transverse relative locality
results for distantly boosted observers in a theory of free particles, which are evidently
as significant as the results for
longitudinal relative locality
established in~\cite{whataboutbob,leeINERTIALlimit}
in the same class of theories for free particles.
We are, as announced, postponing more detailed studies of
transverse relative locality for interacting theories,
of the type introduced in Refs.~\cite{prl,grf2nd}.
But there is another type of manifestation of longitudinal relative locality
for free-particle theories, which we can here reproduce in transverse-relative-locality
version.
These are the results for
longitudinal relative locality  for pure translations
established in Ref.~\cite{kappabob} as a
 peculiar class of coordinate artifacts
 present in a much-studied
 phase-space construction inspired by $\kappa$-Minkowski
noncommutativity~\cite{lukieIW,majidruegg,kpoinap},
in which
one adopts ``$\kappa$-Minkowski Poisson brackets"
for the spacetime coordinates: $\{x_j,t\}= \frac{1}{\kappa} x_j$ (with $\{x_j,x_k\}=0$).
It is easy to verify that there is no transverse relative locality
for pure translations in those $\kappa$-Minkowski-inspired phase-space constructions.
However, in this section we want to show  that some peculiar coordinate
artifacts of transverse relative locality are found under pure translations
within a closely related, still noncommutativity-inspired framework.
For this purpose we introduce the following ``$\rho$-Minkowski"
Poisson brackets for spacetime coordinates\footnote{Note that just
like a 2+1D $\kappa$-Minkowski noncommutativity is linked to
the algebra $hom(2)$ (euclidean-homotheties algebra)
our $\rho$-Minkowski noncommutativity is linked to
the algebra $e(2)$ (euclidean algebra).}
\begin{eqnarray}
& \{x_i,t\}= \rho \epsilon_{ij}x_j \nonumber\\
& \{x_i,x_j\}= 0~ \nonumber
\end{eqnarray}
where $\rho$ is a parameter with dimension of length, and, consistently with
the approach we already adopted in the previous section,
we work at leading order in $\rho$.\\
For the description of space ($P_j$) and time ($\Omega$) translations, the requirement
of enforcing the Jacobi identities\footnote{In particular, it is easy to verify
that instead the standard translations would not satisfy the Jacobi
identities with $\rho$-Minkowski coordinates.}
leads us to 
\begin{eqnarray}
& \{\Omega,t\}=1, \quad \{\Omega,x_i\}=0\nonumber\\
& \{P_i,t\}=0, \quad \{P_i,x_j\}=-(\delta_{ij}+\rho \epsilon_{ij} \Omega)\nonumber\\
& \{P_i,P_j\}=0., \quad \{P_i , \Omega\}=0~.
\end{eqnarray}
The type of transverse-relative-locality
coordinate artifacts we want to characterize in this section
are due to the properties of these translation generators. And to see that
these properties alone suffice to produce
transverse-relative-locality
coordinate artifacts we adopt for this section the standard
on-shell relation
\begin{equation}
\mathcal{C}=\Omega^2 - \vec{P}^2
\label{casimir}
\end{equation}
These ingredients are all that is required for the analysis in this section,
but as a side remark let us observe that the Poisson brackets
we introduced are covariant under classical spatial rotations.
We have in fact that
\begin{eqnarray}
& \{x_i,t\}= \rho \epsilon_{ij}x_j~,~~~\{x_i,x_j\}= 0~ \nonumber\\
&  \Longrightarrow \{x'_i,t'\}= \rho \epsilon_{ij}x'_j~,~~~\{x'_i,x'_j\}= 0 \nonumber
\end{eqnarray}
if
$$t' = t~,~~~
x'_1= x_1 \cos \theta +x_2 \sin \theta ~,~~~x'_2= x_2 \cos \theta -x_1 \sin \theta ~.$$
Moreover, postulating
$$\{R , x_i \} = \epsilon_{ij}x_j~,\{R , t \} = 0~,$$
$$\{R , P_i \} = \epsilon_{ij}P_j~,\{R , \Omega \} = 0~,$$
all Jacobi identities are satisfied.

Within this setup we shall expose transverse relative locality by
analyzing a simultaneous emission of massless particles
in the origin of an observer Alice, as described in the coordinatization
of spacetime by a distant observer Bob.

In preparation for that let us first derive the worldlines
that follow from the undeformed on-shell relation, when analyzed in terms
of our $\rho$-deformed Poisson brackets.
We have that
\begin{eqnarray}
&\dot{t} \equiv \frac{\partial t}{\partial \tau} &
= 2 \Omega \{ \Omega, t \} - 2 P_k \{P_k, t\}= 2\Omega~\\
&\dot{x_i}
\equiv \frac{\partial x_i}{\partial \tau}
&=2 \Omega \{ \Omega, x_i \} - 2 P_k \{P_k, x_i\}
= 2 P_k (\delta_{ik}+\rho \Omega \epsilon_{ik}) ~
\end{eqnarray}
 from which in particular one obtains that for massless particles ($\mathcal{C}=0$)
the worldlines are governed by
\begin{equation}
(x-x_0)_i = (\frac{P_i}{|\vec{P}|} - \rho \epsilon_{ij}P_j) \, (t-t_0)
\end{equation}
Notice that, as a result of the $\rho$-deformed Poisson brackets,
the coordinate velocity depends on momenta in a peculiar way
\begin{equation}
v_i = \frac{P_i}{|\vec{P}|} - \rho \epsilon_{ij}P_j \label{rovel}
\end{equation}
and in particular in order for the coordinate velocity to be directed along
the  $x_1$ direction  actually the momentum must have a small $p_2$ component:
\begin{equation}
\vec{P}=(p, - \rho p^2),\, \Rightarrow \, \vec{v}=(1,0)
\end{equation}
 Still it is easy to check that the speed of massless particles
in this framework is always momentum independent:
\begin{equation}
|\vec{v}|^2 = v_i v_k \delta^{ik} =1 - \rho \frac{1}{|\vec{P}|} \delta^{ik} (\epsilon_{ij} P_j P_k+ \epsilon_{kj} P_j P_i) = 1
\end{equation}

We are now ready to consider the emission at Alice of two massless
particles, a soft particle of momentum $p^{(s)}$ (with $\rho p^{(s)} \simeq 0$)
and a hard particle of momentum $p^{(h)}$ (such $\rho p^{(h)}$ cannot be neglected).
And let us further restrict our focus to the case in which, according to Alice's
coordinates, the two massless
particles are emitted toward the $x_1$ axis.
In light of the results derived above
it is evident that according to Alice the two wordlines are coincident:
\begin{eqnarray}
x^{(s)A}_1(t^A) = t^{A}~,~~~
x^{(s)A}_2(t^A) = 0.
\label{rhowl1}
\end{eqnarray}
\begin{eqnarray}
x^{(h)A}_1(t^A) = t^{A}~,~~~
x^{(h)A}_2(t^A) = 0.
\label{rhowl2}
\end{eqnarray}

We can now establish how the peculiar properties of our $\rho$-deformed
translations affect the way in which a distant observer Bob describes these
two worldlines.
In general for our $\rho$-deformed
translations
we have
\begin{eqnarray}
\mathcal{T}_{a_t , a_1 , a_2 } \rhd t & = & t - a_t \{\Omega,t\} + a_j \{P_j,t\} = t - a_t,\nonumber\\
\mathcal{T}_{a_t , a_1 , a_2 } \rhd x_i & = & x_i - a_t \{\Omega,x_i\} + a_j \{P_j,x_i\} = x_i - a_i + \rho \Omega \epsilon_{ij} a_j.
\end{eqnarray}
We are again interested in the description of the two worldlines
given by an observer Bob, at rest with respect to Alice, and such that
the soft massless particle reaches Bob in his spacetime origin.
So the translation parameters of our interest are $ a_t = a,~a_1 = a,~a_2 = 0$.
For this choice of translation parameters we find that the two worldlines
are, according to Bob,
\begin{eqnarray}
x^{(s)B}_1(t^B) = t^B~,~~~
x^{(s)B}_2(t^B) = 0.
\end{eqnarray}
\begin{eqnarray}
x^{(h)B}_1(t^B) = t^B~,~~~
x^{(h)B}_2(t^B) =  -\rho a p^{(h)}.
\end{eqnarray}

As shown in Figure 3, these results confirm that with $\rho$-Minkowski coordinates
one finds transverse relative locality which is completely analogous to
the longitudinal relative locality found with $\kappa$-Minkowski coordinates
in Ref.~\cite{kappabob}. In particular we have seen that a coincident emission
at Alice, is described as a pair of non-coincident emission events by distant
observer Bob, and the lack of coincidence is seen by Bob in the direction orthogonal
to the direction of the translation connecting Alice to Bob.

\begin{figure}[h!]
\begin{center}
\includegraphics[scale=0.24]{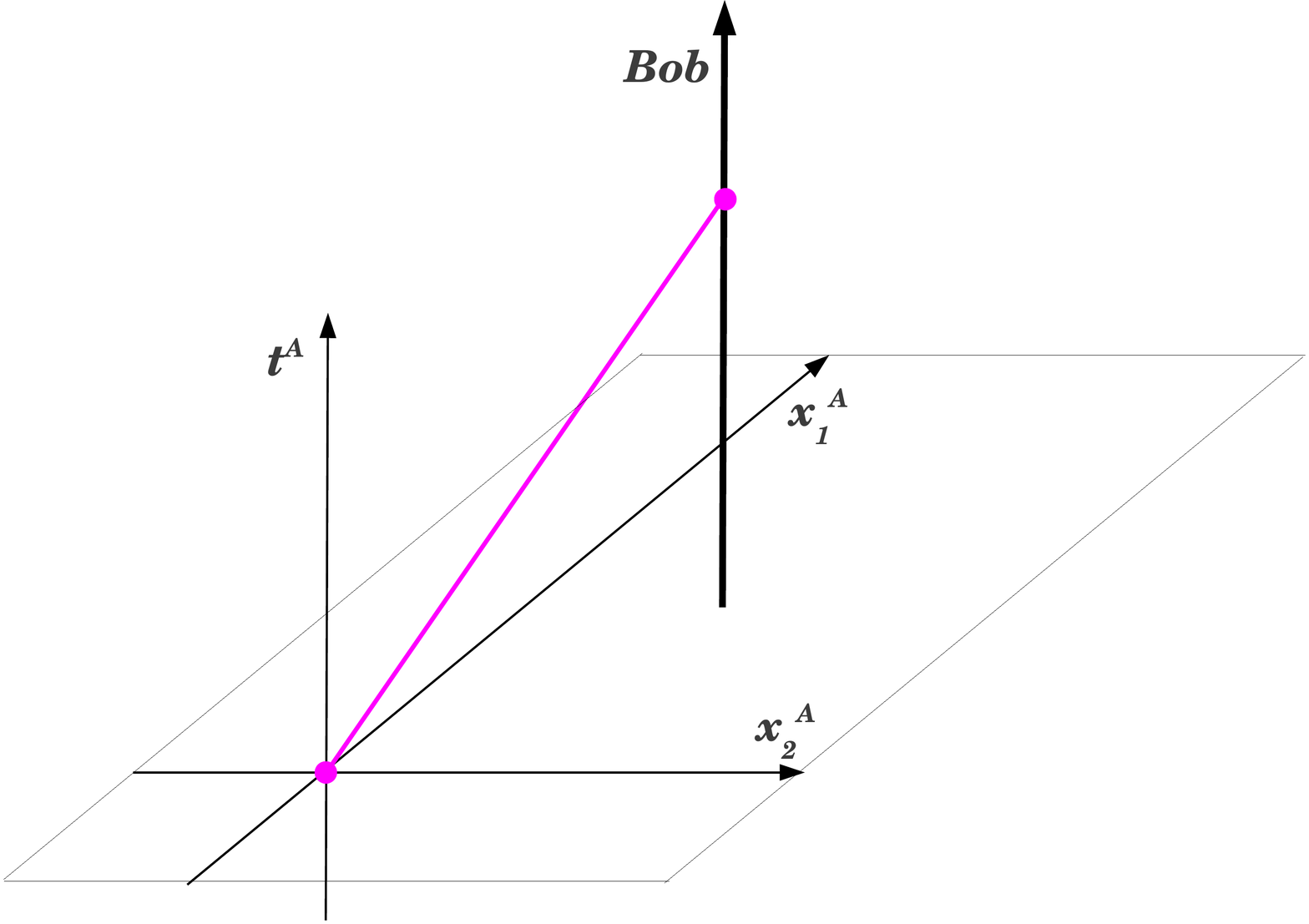}
\includegraphics[scale=0.24]{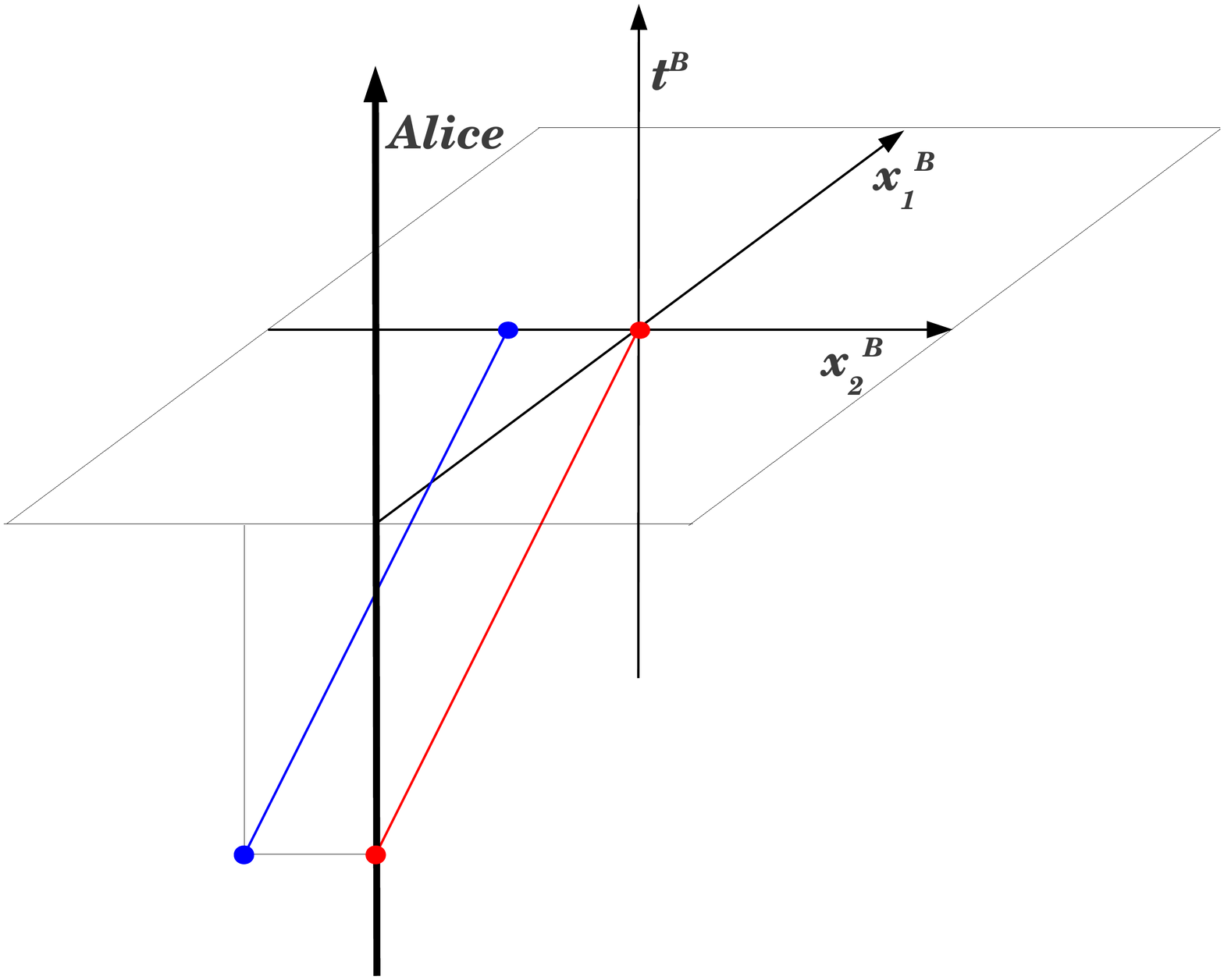}
\includegraphics[scale=0.22]{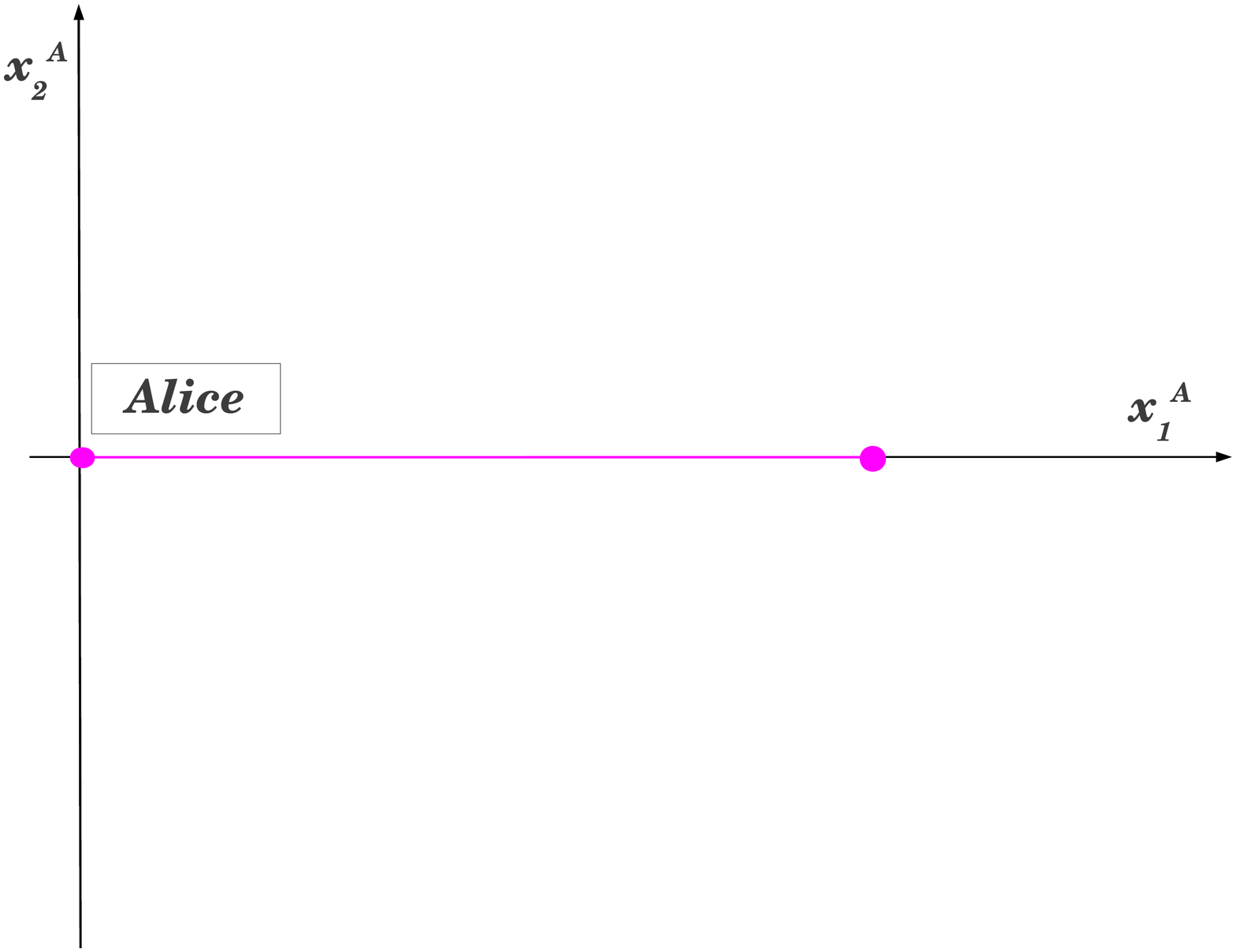}
\includegraphics[scale=0.22]{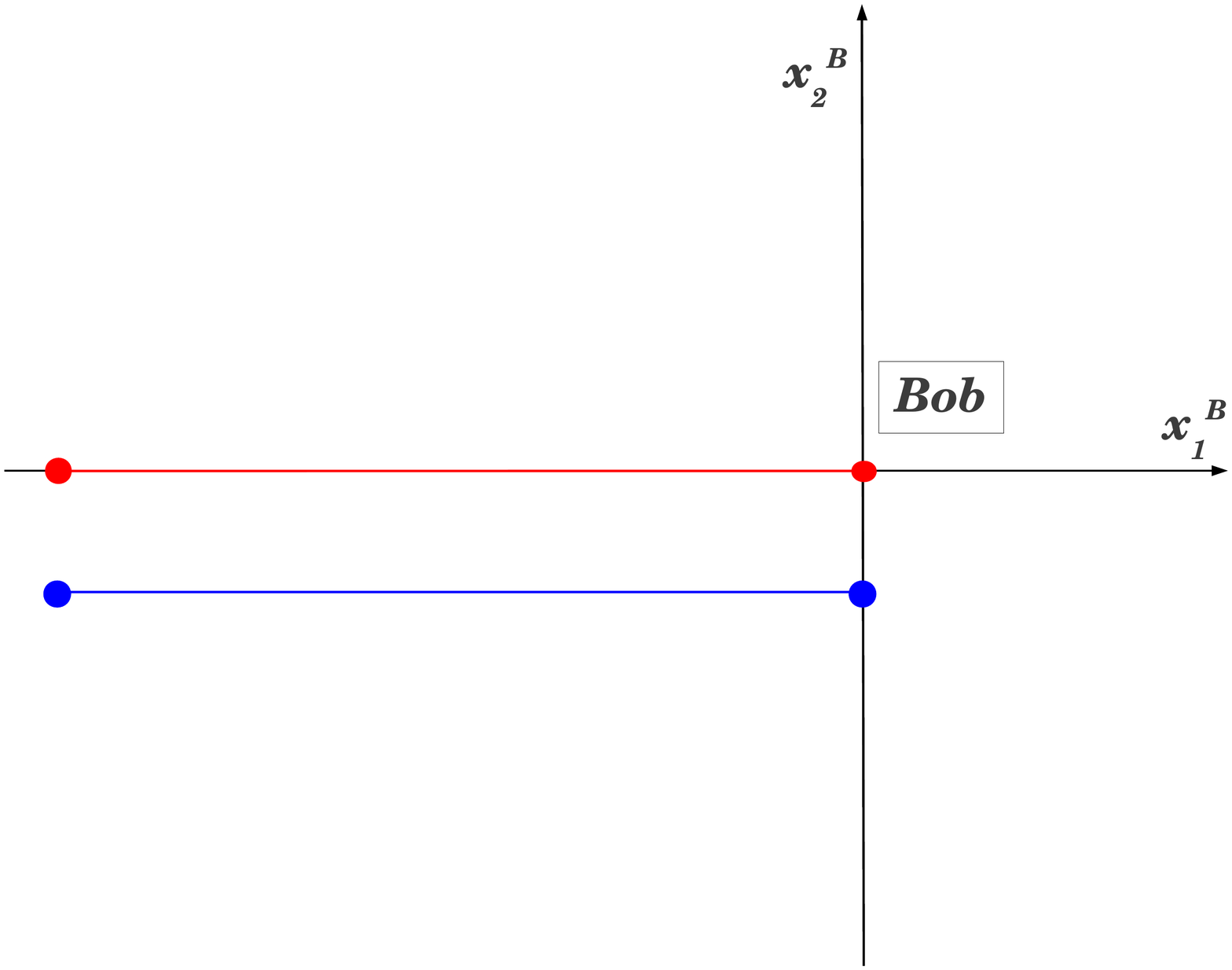}
\caption{In this case Alice (left panels) and Bob (right panels)
use $\rho$-Minkowski coordinates.
We show
3D worldlines
(top panels) and their 2D
spatial projection (bottom panels) for a soft and a hard (respectively red and blue; violet when coincident) massless particles, emitted simultaneously at Alice
toward Bob. In Bob's coordinates the two worldlines do not coincide and
 there is some transverse relative locality: the coincidence of emission events
witnessed by nearby observer Alice, is not present in the coordinatization by distant observer Bob, and the difference between emission points in Bob coordinates is purely along the $x_2$
axis, orthogonal to the direction ($x_1$) of the translation transformation
from Alice to Bob.}
\end{center}
\end{figure}

\section{Closing remarks}
We feel we have here accomplished the main task we had set for ourselves,
by showing
that transverse relative locality is an aspect
of relative locality that cannot be dismissed, and actually may deserve
as much attention as longitudinal relative locality.
This was established here within the confines of relative-locality theories
of free particles,  but it is hard to imagine that for
interacting relative-locality particles, which are described
within the framework of Refs.~\cite{prl,grf2nd},
the ``balance of power" between longitudinal and transverse relative
locality could be significantly shifted.

Our results on transverse relative locality and dual-gravity lensing
in theories of free particles may also provide some guidance
for future more detailed analyses of such features
within the framework of Refs.~\cite{prl,grf2nd}
for interacting particles. These should
also take as starting point the preliminary results on dual-gravity lensing
reported in parts of Ref.~\cite{leelaurentGRB}. In particular,
the feature of dual-gravity lensing exposed in
Ref.~\cite{leelaurentGRB} was proportional to the sum of the energies(/momenta)
of the two particles whose wordlines were experiencing lensing.
It was already clear from Ref.~\cite{leelaurentGRB} that this result of dependence
on the sum of energies had only been checked within a very specific setup
for the derivation, including definite choices among the many possible
chains of interactions that could be considered in the interacting-particle framework.
The fact that here,  within the limitations of a theory of free particles,
we found some dual-gravity lensing proportional to the difference
of the energies(/momenta)
of the two particles whose wordlines experience lensing can provide encouragement
for the search of other chains of interactions,
in which the difference of energies governs the dual-gravity lensing.

The scopes of our analysis were still too limited for allowing speculations
about phenomenology, but it is nonetheless noteworthy from that perspective
that in Section II we found transverse-relative-locality
effects of exactly the same magnitude of the effects of longitudinal
relative locality previously found in Ref.~\cite{whataboutbob},
and those are (at least indirectly) testable~\cite{whataboutbob,magicELLISnew,unoEdue}, even if $\ell$ is of the order of
the Planck length.

The brief exploration, in Sec.~III, of transverse relative locality when observers
adopt our ``$\rho$-Minkowski coordinates" must be viewed in exactly the same spirit as
the analogous results obtained in Ref.~\cite{kappabob} for observers
adopting ``$\kappa$-Minkowski coordinates". In theories of classical particles
these results may at best clarify possible confusion arising with the
use of such non-standard coordinates (the confusion addressed in Ref.~\cite{kappabob}
being a particularly strong example, since it had obstructed progress in
a relevant research area for more than a decade). But  we feel such
preliminary studies of classical theories with non-standard coordinates
should have as ultimate goal
the development of suitable quantum versions. And just
like the use of ``$\kappa$-Minkowski coordinates" in classical theories does
prepare one's intuition for studies of the ``$\kappa$-Minkowski
non-commutative spacetime"~\cite{majidruegg,kpoinap}, we expect that
our preliminary observations on ``$\rho$-Minkowski coordinates" might
set the stage for intriguing studies of ``$\rho$-Minkowski noncommutativity".

Finally, we feel that the results we here reported should have some influence
on future developments of the ``doubly-special relativity"
research programme~\cite{gacdsr1,kowadsr,leedsrPRL,dsrnature,leedsrPRD,jurekDSRnew}.
Hundreds of papers have been devoted over the last decade to
doubly-special-relativity results formulated exclusively
in momentum space (see, {\it e.g.}, Refs.~\cite{gacdsr1,kowadsr,leedsrPRL,dsrnature,leedsrPRD,jurekDSRnew}
and references therein). Only recently some spacetime aspects of
doubly-special relativity were satisfactorily analyzed
in a handful of studies, which were however confined to
essentially 1+1-dimensional analyses, and led to
 some of the first results
on longitudinal relative locality~\cite{whataboutbob,leeINERTIALlimit,arzkowaRelLoc}.
We here reported, in Section II, an analysis in which the presence
of more than one spatial dimension in a
 doubly-special-relativity framework plays a nontrivial role,
and we feel this could now set the new standard for studies
attempting to advance doubly-special-relativity research.\\
$~$\\
$~$ $~~~~~~~~~~~~~~~~~~~~~~$ \textbf{ACKNOWLEDGEMENTS}\\
We are grateful to Michele Arzano for valuable feedback on a first draft of
this manuscript. Our work also benefitted from discussions of various aspects
of relative locality with Laurent Freidel, Giulia Gubitosi, Jerzy Kowalski-Glikman, 
Flavio Mercati, Giacomo Rosati and Lee Smolin.


\newpage

\begin{thebibliography}{50}

\bibitem{prl}
  G.~Amelino-Camelia, L.~Freidel, J.~Kowalski-Glikman
L.~Smolin,
 arXiv:1101.0931.

\bibitem{grf2nd}   G.~Amelino-Camelia, L.~Freidel, J.~Kowalski-Glikman
L.~Smolin,
 arXiv:1106.0313
[Awarded 2nd prize in the 2011 competition hosted by the Gravity Research Foundation].


\bibitem{whataboutbob}
  G.~Amelino-Camelia, M.~Matassa, F.~Mercati and G.~Rosati,
  arXiv:1006.2126,
  Phys.~Rev.~Lett.~\textbf{106}, 071301 (2011).

\bibitem{leeINERTIALlimit}
L.~Smolin, arXiv:1007.0718.

\bibitem{arzkowaRelLoc} M.~Arzano, J.~Kowalski-Glikman,
 arXiv:1008.2962,
 Class.~Quant.~Grav.~\textbf{28}:105009 (2011).

\bibitem{gacdsr1} G.~Amelino-Camelia, arXiv:gr-qc/0012051,
Int.~J.~Mod.~Phys.~\textbf{D11} (2002) 35;
 arXiv:hep-th/0012238,
 Phys.~Lett.~\textbf{B510} (2001) 255.

\bibitem{kowadsr} J.~Kowalski-Glikman,
 arXiv:hep-th/0102098,
 Phys.~Lett.~\textbf{A286} (2001) 391.

\bibitem{leedsrPRL}
  J.~Magueijo, L.~Smolin,
 arXiv:hep-th/0112090,
 Phys.~Rev.~Lett.~{\bf 88} (2002)  190403.

\bibitem{dsrnature}
G.~Amelino-Camelia,
arXiv:gr-qc/0207049,
 Nature \textbf{418} (2002) 34.

 \bibitem{leedsrPRD}
  J.~Magueijo, L.~Smolin,
 arXiv:gr-qc/0207085,
 Phys.~Rev.~\textbf{D67} (2003) 044017.

\bibitem{jurekDSRnew} J.~Kowalski-Glikman, S.~Nowak,
 arXiv:hep-th/0204245,
 Int.~J.~Mod.~Phys.~\textbf{D12} (2003) 299.

\bibitem{leelaurentGRB}
 L.~Freidel, L.~Smolin,
 arXiv:1103.5626.

\bibitem{soccerball}
  G.~Amelino-Camelia, L.~Freidel, J.~Kowalski-Glikman,
L.~Smolin,
 arXiv:1104.2019.

\bibitem{flaviogiuliaRL}
G.~Gubitosi and F.~Mercati,
 arXiv:1106.5710.

\bibitem{flaviojoseRL}
J.M.~Carmona, J.L.~Cortes, D.~Mazon and F.~Mercati,
 arXiv:1107.0939

\bibitem{anatomy}
  G.~Amelino-Camelia, M.~Arzano, J.~Kowalski-Glikman,
 G.~Rosati and G.~Trevisan,
 arXiv:1107.1724.


\bibitem{lukieIW} J.~Lukierski, H.~Ruegg, A.~Nowicki
and V.N.~Tolstoi, Phys.~Lett.~\textbf{B264} (1991) 331;
J.~Lukierski, A.~Nowicki and H.~Ruegg,
Phys.~Lett.~\textbf{B293} (1992) 344.

\bibitem{majidruegg} S.~Majid and H.~Ruegg:
{\ Phys.~Lett.}~B~\textbf{334}, 348 (1994).

\bibitem{kpoinap} J.~Lukierski, H.~Ruegg and W.J.~Zakrzewski:
{\ Ann.~Phys.}~\textbf{243}, 90 (1995).

\bibitem{kappabob}
  G.~Amelino-Camelia, N.~Loret and G.~Rosati,
  arXiv:1102.4637,
  Phys.~Lett.~\textbf{B700} (2011) 150.

\bibitem{jurekvelISOne} M.~Daszkiewicz, K.~Imilkowska
and J.~Kowalski-Glikman,
 arXiv:hep-th/0304027,
 Phys.~Lett.~{\bf A323} (2004) 345.

\bibitem{mignemi}
S.~Mignemi,
 arXiv:gr-qc/0304029,
 Phys.~Rev.~\textbf{D68}, 065029 (2003).

\bibitem{gacMandaniciDANDREA}
  G.~Amelino-Camelia, F.~D'Andrea and G.~Mandanici,
 arXiv:hep-th/0211022,
  JCAP~{\bf 0309} (2003) 006.

\bibitem{mignemiVEL} S.~Mignemi, arXiv:hep-th/0302065,
Phys.~Lett.~{\bf A316} (2003) 173.

\bibitem{ghoshVEL}
S.~Ghosh and P.~Pal, arXiv:hep-th/0702159, Phys.~Rev.~{\bf D75} (2007) 105021.

\bibitem{magicELLISnew}
J.~Ellis, N.E.~Mavromatos and D.V.~Nanopoulos,
 arXiv:0901.4052, 
 Phys.~Lett.~{\bf B 674} (2009) 83.

\bibitem{unoEdue}
G. Amelino-Camelia and L. Smolin,
 arXiv:0906.3731, 
 Phys.~Rev.~{\bf D80} (2009) 084017.

\end{thebibliography}
\end{document}